\documentclass[aps,prb,groupedaddress,twocolumn,showpacs]{revtex4}
\usepackage{graphicx}

\begin{document}

\title{Critical behavior and driven Monte Carlo dynamics
of the XY spin glass in the phase representation}

\author{Enzo Granato}

\address{Laborat\'orio Associado de Sensores e Materiais, \\
Instituto Nacional de Pesquisas Espaciais, \\
12245-970 S\~ao Jos\'e dos Campos, SP Brazil}

\begin{abstract}
A driven Monte Carlo dynamics is introduced to study resistivity
scaling in XY-type models in the phase representation. The method
is used to study the phase transition of the three-dimensional XY
spin glass with a Gaussian coupling distribution. We find a
phase-coherence transition at finite temperature in good agreement
with recent equilibrium Monte Carlo simulations which shows a
single (spin and chiral) glass transition. Estimates of the static
and dynamic critical exponents indicate that the critical behavior
is in the same universality class as the the model with a bimodal
coupling distribution. Relevance of these results for
$\pi$-junction superconductors is also discussed.

\end{abstract}


\maketitle

Phase transition in the three-dimensional (3d) XY spin glass with
short-range interactions has been a challenging problem. The
additional chiral order parameter \cite{villain}, with Ising-like
symmetry, arising from frustration effects competes with spin
glass ordering due to freezing of the two component spins. Earlier
work showed that while the spin glass transition temperature
vanishes, the chirality orders at a finite temperature
\cite{kawa87}. Evidence of such behavior has been provided both
for models where the coupling  between XY spins has a bimodal
distribution ( $+1$ and $-1$ with equal probability) and for
models with a Gaussian coupling distribution. However, recent
numerical work based on equilibrium and dynamic simulations has
questioned this decoupled scenario with some conflicting results.

Like other XY type models \cite{fisher}, the XY spin glass can
also be regarded as a model for phase-coherence in
superconductors. In particular, it is currently being used as a
model for granular superconductors containing $\pi$ junctions
\cite{sigrist,kawamura,kawa01,eg,eg01,eg04}, as in high-$T_c$
superconductor materials with d-wave symmetry \cite{sigrist}. In
this case, a $\pi$ junction corresponds to an antiferromagnetic
coupling between the XY spins while the orientational angle of the
spins represents the phase of the local superconductor order
parameter. This superconductor analog allows the use of the
electrical resistivity as a very useful dynamic quantity that can
characterize spin-glass ordering in the XY spin glass. In fact,
the linear resistivity is a direct measure of phase stiffness and
therefore phase coherence in the superconductor, which is
equivalent to long-range order in the spin variables.  Thus, the
resistivity behavior can be used to study numerically the phase
transition in the XY spin glass.

Earlier Monte Carlo (MC) simulations of resistivity behavior of
the XY spin glass with a bimodal coupling distribution in the
vortex representation \cite{wengel}, showed evidence of a
resistive transition at finite temperature. However, this result
was interpreted as an indication of the chiral-glass transition.
With a different interpretation \cite{eg}, it was argued that this
resistive transition should instead be attributed to the
spin-glass ordering. The possibility of spin-glass ordering at
finite temperatures was also supported by calculations of the spin
stiffness exponent in the ground state \cite{grempel,kost},
showing that the lower-critical dimension is below $3$, which
implies that a phase-coherence transition at finite-temperature is
possible in three dimensions. Later on, resistivity calculations
in the phase representation using Langevin dynamics
\cite{eg01,eg04} confirmed the occurrence of a phase-coherence
transition at finite temperature and suggested a single transition
scenario where chiral and phase variables order simultaneously.

Recent equilibrium MC simulations by Kawamura and Li for the XY
spin glass with a bimodal coupling distribution \cite{kawa03},
provided an improved estimate of the chiral-glass transition
temperature, which turned out to be consistent with the transition
temperature as obtained from the resistivity scaling \cite{eg04},
but still conclude for the absence of phase coherence within a
spin-chirality decoupling  scenario. However, very recently, MC
calculations of the chiral and spin correlation lengths by Lee and
Young for the model with Gaussian couplings \cite{ly} show that
there is indeed a transition at finite temperature at which both
spin and chirality order, supporting the single transition
scenario. These conflicting results raise the immediate question
of universality class of the phase transition in the two models.
In addition, since the resistivity scaling approach indicates a
phase-coherence transition for the bimodal coupling model at the
chiral-glass transition, it should also be of particular interest
to find out if the gives consistent results for the Gaussian
coupling model.

In this work, we study the critical behavior of the 3d XY spin
glass with Gaussian couplings by a scaling analysis of the
nonlinear resistivity obtained by MC methods.  First, we introduce
a driven MC dynamics for XY-type models in the phase
representation. The alternative MC dynamics in the vortex
representation \cite{wengel}, which has been used for the model
with bimodal couplings, is not useful for the Gaussian model
because in this case the coupling magnitude is not uniform.  The
main advantage of the driven MC method in the phase representation
compared with standard Langevin dynamics simulations
\cite{eg01,eg04} is that much longer time scales can be accessed,
insuring that the long-time behavior is probed at the lowest
temperatures and current densities. In addition, the driven MC
dynamics is particularly useful in models for which Langevin phase
dynamics or MC vortex dynamics are not available \cite{xyising}.

From a scaling analysis of the resistivity data for the 3d XY spin
glass, we find a phase-coherence transition at finite temperature
and the corresponding thermal and dynamic critical exponents are
determined. The results are in good agreement with recent
equilibrium MC simulations for the model with Gaussian couplings
\cite{ly}. Moreover, comparing the static and dynamic critical
exponents obtained for the Gaussian  model with recent results for
the bimodal coupling model using the same analysis
\cite{eg01,eg04} suggests a common universality class. Altogether,
these results strongly support the single transition scenario
\cite{eg01} where chiral and phase variables order simultaneously
in both models.

We consider the XY-spin glass driven by an external perturbation,
described by the Hamiltonian
\begin{equation}
H=-\sum_{<ij>}J_{ij}\cos(\theta_i - \theta_j) -J \sum_i(\theta_i -
\theta_{i+x}) \label{xyspin}
\end{equation}
The first term gives the nearest-neighbor coupling energy, where
$J_{ij}$ are quenched random variables taken from  the Gaussian
distribution with zero mean and unit variance
\begin{equation}
P(J_{ij})= \frac{1}{\sqrt{2\pi}} \exp(-J_{ij}^2 /2).
\end{equation}
The second term in Eq. \ref{xyspin} represents the effects of an
external perturbation, applied in the $x$-direction for
convenience, coupled to the bond phase difference $\theta_i -
\theta_{i+x}$. When regarded as a model for granular
superconductors with $\pi$ junctions \cite{kawa03,eg04}, the
random distributed negative signs of the bond variable $J_{ij}$
correspond to Josephson junctions with a phase shift of $\pi$ and
the perturbation is equivalent to a driving current density $J$
applied to the superconductor. When $J \ne 0$, the total energy is
unbounded and the system is out of equilibrium. The lower energy
minima occur at increasing phase differences $ \theta_i -
\theta_{i+x}$ as a function of time, leading to a net phase
slippage rate proportional to  $<d (\theta_i - \theta_{i+x})/
dt>$, which can be taken as a measure of the voltage $V$ (in
arbitrary units) in a model of superconductors.

To study the nonequilibrium behavior generated by the driving
current density $J$ in Eq. \ref{xyspin}, we use a driven MC
dynamics method. The time dependence is obtained by identifying
the MC time as the real time $t$ and we set the unit of time
$dt=1$, corresponding to a complete MC pass through the lattice.
Periodic (fluctuating twist) boundary conditions are used
\cite{saslow} in cubic systems of linear size $L$. This boundary
condition adds new dynamical variables, $u_\alpha$ ($\alpha =x,y$
and $z$), corresponding to a uniform phase twist between
nearest-neighbor sites along the principal axis directions $\hat
x, \hat y$ and $\hat z$. A MC step consists of an attempt to
change the local phase $\theta_i$  and the phase twists $u_\alpha$
by fixed amounts, using the Metropolis algorithm. If the change in
energy is $\Delta H$, the trial move is accepted with probability
$ min\{1,\exp(-\Delta H/kT)\}$. The external current density $J$
in Eq. \ref{xyspin} biases these changes, leading to a net voltage
(phase slippage rate) across the system, given by
\begin{equation}
 V = \frac{1}{L^{2}}\frac{d}{dt}  \sum_{j,k=1}^L(\theta_{1,j,k} - \theta_{L,j,k}-u_xL),
\end{equation}
Using this procedure, the electric field $E=V/L$ and nonlinear
resistivity $\rho = E/J$ can be obtained as a function of the
driving current density $J$ for different temperatures $T$. The
main advantage of this MC method compared with standard Langevin
dynamics \cite{eg,eg01} for models in the phase representation is
that in principle much longer time scales can be accessed. In the
latter method, the maximum time step is limited by numerical
instabilities when integrating the Langevin differential equations
whereas in the former, trial moves are possible which would
correspond to very large time steps in the numerical integration.

For the simulations of  resistivity behavior described below,
first MC calculations are performed with $J=0$ (zero current bias)
to obtain the equilibrium state which is then used as initial
state for the driven MC dynamics with $J \ne 0$. Extensive
calculations were performed, using typically $10^7$ MC steps for
the equilibration and driven MC dynamics with $10$ to $20$
different realizations of the $J_{ij}$ disorder distribution for
low values of $J$ in system sizes ranging from $L=4$ to $L=12$.

To extract the critical behavior from the numerical results of the
nonlinear resistivity $\rho$ we need a scaling theory for the
resistive behavior near a second-order phase transition. A
detailed scaling theory has been described in the context of the
current-voltage characteristics of vortex-glass models
\cite{fisher} but it can be directly applied here. If a
phase-coherence transition occurs at nonzero temperature $T_c$,
then measurable quantities scale with the diverging  correlation
length $\xi \propto |T/T_c -1|^{-\nu}$ and relaxation time $\tau
\propto \xi^z$, where $\nu$ and $z$ are the thermal and dynamical
critical exponents, respectively. The nonlinear resistivity $\rho$
should then satisfy the scaling form \cite{fisher}
\begin{equation}
T \rho \ \xi^{\ z-1}= g_\pm(\frac{J \xi^{2}}{T}) \label{scaltc}
\end{equation}
where $g(x)$ is a scaling function. The $+$ and $-$ signs
correspond to $T>T_c$ and $T<T_c$, respectively. If the numerical
data satisfy such scaling form for different temperatures and
driving currents, then the critical temperature and critical
exponents of the underlying equilibrium transition at $J=0$ can be
estimated from the best data collapse. However, for a reliable
estimate, the data should also satisfy the expected finite-size
behavior in smaller system sizes. Finite-size effects are
particularly important sufficiently close to $T_c$ when the
correlation length $\xi$ approaches the system size $L$. In
particular, at $T_c$, the correlation length will be cut off by
the system size in any finite system and the nonlinear resistivity
should then satisfy a scaling form as in Eq. \ref{scaltc} with
$\xi = L$,
\begin{equation}
T_c \ \rho \  L^{\ z-1}= g(\frac{J L^{2}}{T_c}) \label{scaltcL}
\end{equation}
Away from $T_c$, the scaling function in Eq. \ref{scaltc} will
also depend on the dimensionless ratio \cite{fisher,wengel}
$L/\xi$ as $g(J\xi^{2}/T,L/\xi)$. To simplify the analysis, we
consider resistivity data at current densities such that
$J\xi^{2}/T=$ is constant. Then, the scaling form depends only on
a single variable and the resistivity should satisfy the
finite-size scaling form
\begin{equation}
 T \rho \ L^{z-1}= \tilde{g}(L^{1/\nu}(T/T_c-1))  \label{scalL}
 \end{equation}

\begin{figure}
\includegraphics[bb= 1cm  2.5cm  19cm   16.5cm, width=7.5 cm]{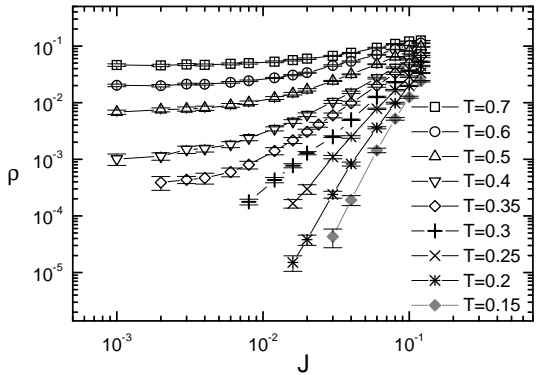}
\caption{Nonlinear resistivity $\rho$ for  different temperatures
$T$, for a system size $L=12$. }
\end{figure}

We now describe the results of the scaling analysis of the
resistivity behavior obtained from the driven MC dynamics.  The
nonlinear resistivity $\rho$ as a function of current density and
temperature is shown in Figs. 1 for a large system size $L=12$.
The behavior is consistent with a phase-coherence transition at an
apparent critical temperature in the range $T_c \sim 0.25 -0.35$.
At higher temperatures, the linear resistivity
$\rho_L=\lim_{J\rightarrow 0} E/J$ is finite while  at lower
temperatures, it extrapolates to very low values. The actual
limiting values of the resistivity at low currents can not be
determined accurately with the available computer time.  However,
if one assumes a continuous equilibrium transition at a finite
temperature with $J=0$, then the nonlinear resistivity should
satisfy the scaling form of Eqs. \ref{scaltc} and the critical
temperature and exponents can then be obtained from the data
collapse. Such scaling plot is shown in Fig. 2, obtained by
adjusting the unknown parameters. We now show that theses
estimates, using the largest system size, are reliable by
verifying that they satisfy the expected finite-size behavior
using smaller system sizes. In fact, as shown in Fig. 3, the
nonlinear resistivity satisfy the finite-size scaling form of Eq.
\ref{scaltcL} for different system sizes at the estimated
$T_c=0.335$. Also, away from $T_c$, the nonlinear resistivity
calculated at different temperatures and system sizes satisfy the
scaling form of Eq. \ref{scalL} as shown in Fig. 4. From this
scaling analysis we obtain the estimate of critical temperature
and critical exponents $T_c=0.335(15)$, $z=4.5(3)$ and $\nu =
1.2(2)$. It should be emphasized here that this result is obtained
by requiring that $T_c$, $z$ and $\nu$ satisfy, not only the
scaling form of Eq. \ref{scaltc} for the large system
\cite{holzer}, but at same time also the finite-size scaling forms
of Eqs. \ref{scaltcL} and \ref{scalL} for smaller systems.

\begin{figure}
\includegraphics[bb= 1cm  2.5cm  19cm   16.5cm, width=7.5 cm]{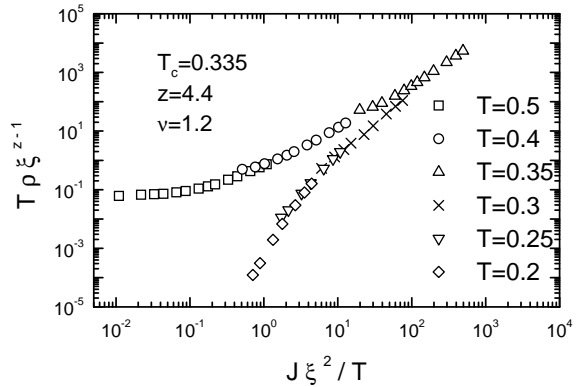}
\caption{ Scaling plot of the data from Fig. 1 near the
transition, with $\xi \propto |T/T_c - 1|^{-\nu}$. }

\end{figure}

Our estimate of the phase-coherence critical temperature and
exponent for the model with Gaussian couplings is particularly
interesting in view of the recent equilibrium MC simulations
\cite{ly} which show evidence of spin glass transition at finite
temperature and that chirality and spin variables order
simultaneously. Our results provide further support for the
estimate of the spin-glass transition at $T_c =  0.34(2)$, as
first reported in that work. Moreover, our estimate of the static
and dynamic critical exponents agree with those obtained for the
model with bimodal coupling distribution using similar analysis
\cite{eg01,eg04} ( $\nu =1.2(2)$, $z=4.4(3)$ ), suggesting a
common static and dynamic universality class. If indeed the
critical behavior is the same for both models then a single
transition should also be observed in the latter model, despite
the conclusions from other MC simulations \cite{kawa03} that find
a spin-chirality decoupling scenario.

\begin{figure}
\includegraphics[bb= 1cm  2.5cm  19cm   16.5cm, width=7.5 cm]{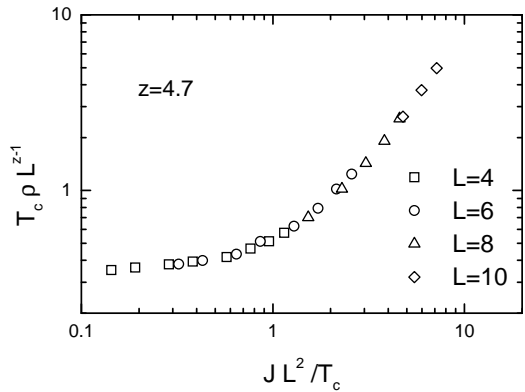}
\caption{Finite-size scaling plot of the nonlinear resistivity at
$T_c=0.335 $.}
\end{figure}

Our results for the resistive transition in the Gaussian coupling
model are also relevant for understanding the behavior of granular
superconductors with $\pi$ junctions
\cite{kawamura,kawa01,eg01,eg04}. In particular, some measurements
in high-$T_c$ superconductor materials \cite{yamao}, and numerical
simulations \cite{lid}, showing a power-law behavior for the
nonlinear contribution to the resistivity near the onset of the
paramagnetic Meissner effect, have been interpreted as resulting
from a chiral-glass transition with no phase coherence, based on
earlier results for the XY-spin glass model with a bimodal
coupling distribution. In view of the results for resistivity
scaling, an alternative interpretation is possible \cite{eg04}
where the observed behavior is a consequence of the underlying
phase-coherence transition and the power-law exponent $\alpha$ is
determined by the dynamic critical exponent $z$ as
$\alpha=(5-z)/2$. In this regard, the results for the Gaussian
model considered here have two important implications. First, this
model is a more realistic description of the superconductor
material since in a granular sample the Josephson couplings
between grains will have random magnitudes, as well as random
signs due to the $\pi$ phase shifts. Secondly, our results show
that a phase-coherence transition takes place at finite
temperature  and the common universality class we find for the
resistive transition in these models shows that the proposed
\cite{eg04} superconducting chiral transition and the power-law
exponent $\alpha$, do not depend on details of the coupling
distribution.

\begin{figure}
\includegraphics[bb= 1cm  2.5cm  19cm   16.5cm, width=7.5 cm]{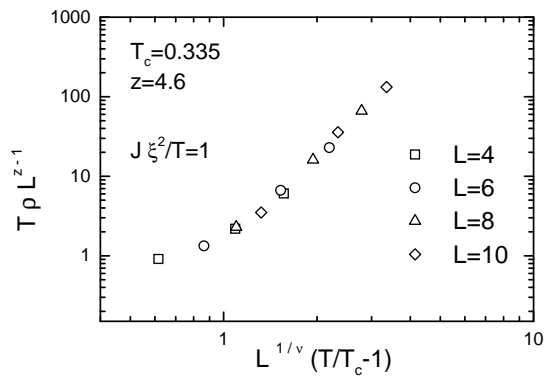}
\caption{Finite-size scaling plot near $T_c$ using current
densities such that $J\xi^{2}/T=1$, a constant value.}
\end{figure}

In summary, we have introduced a driven MC dynamics method to
determine the resistivity behavior of XY type models in the phase
representation. The method is used to study resistivity scaling
and the phase transition in the 3d XY spin glass. From the scaling
analysis we find clear evidence of a phase-coherence transition at
finite temperature. The critical temperature and exponents are in
good agreement with recent equilibrium MC simulations with a
Gaussian coupling distributions \cite{ly} and suggest that the
critical behavior is in the same universality class as the the
model with a bimodal coupling distribution \cite{eg04}. For
$\pi$-junction superconductors, the results demonstrate that the
superconducting chiral-glass transition and the numerical value of
power-law exponent $\alpha$, are not sensitive to the  details of
the coupling distribution.

\bigskip \noindent This work was supported by FAPESP (grant no.
03/00541-0).

\end{document}